\documentclass[preprint2]{aastex}

\slugcomment{}

\shorttitle{Optical studies of V4332 Sgr}
\shortauthors{AAAA et al.}

\begin{document}


\title{Optical studies of V4332 Sagittarii - detection of unusually strong
 KI and NaI lines in emission}


\author{Dipankar P.K. Banerjee}
\affil{Physical Research Laboratory, Navrangpura,  Ahmedabad 
Gujarat 380009,India}
\email{orion@prl.ernet.in}
\and
\author{Nagarhalli M. Ashok}
\affil{Physical Research Laboratory, Navrangpura,  Ahmedabad 
Gujarat 380009,India}
\email{ashok@prl.ernet.in}


\begin{abstract}
        
               We present optical observations of the enigmatic nova-like
        variable V4332 Sgr. The importance of this object should not be
        understated since it is considered to be the possible prototype of
        a new class of eruptive variables.  These objects have been the
        subject of considerable studies at present primarily because of
        the spectacular eruption of V838 Mon - another member of this
        class - recently in 2002.  The cause of the outburst in such
        objects is not well understood. Our recent work has shown striking
        changes in the near-IR spectrum of V4332 Sgr since its 1994
        outburst. The optical spectrum presented here confirms that V4332
        Sgr is indeed an unusual and extremely interesting object. This
        spectrum, the first to be taken after a hiatus of nearly 10 years
        after the outburst, shows several lines in emission but is
        dominated by exceptionally strong emission in the resonance
        doublet of KI at 7665 and 7699{\AA} and to a slightly lesser
        strength in the unresolved NaI doublet at 5890 and 5896{\AA}.
        The KI lines are shown to be optically thick and
        considerably broadened.  We investigate the site of
        origin of the KI and NaI emission.  Considering the strength of
        the alkali metal lines - seen at similar strength only in L and T type
        dwarfs (though in absorption) - we discuss whether the outburst of
        V4332 Sgr was an explosion on a L or T type dwarf.  However $BVRI$
        photometry does not support such a scenario but rather shows the
        central object of V4332 Sgr to be a M-type star with a temperature
        of 3250K.

\end{abstract}




\section{Introduction}
      
               V4332 Sagittarii (V4332 Sgr) erupted in 1994 in a nova-like
        explosion which was recognized to be unusual (Martini et al. 1999).
        The object showed a rapid post-outburst evolution to a cool M
        giant/supergiant which was uncharacteristic of a classical nova
        outburst. The light curve of the object (Martini et al. 1999) showed
        a slow rise to a maximum visual magnitude of $\sim$ 8.5. This was
        followed by a fast decline with the decay time 
        for 2 and 3 magnitudes being 8$\pm$1 and 12$\pm$1 days respectively.
        Current interest in V4332 Sgr was resurrected by the January
        2002 eruption of  V838 Mon which showed the outburst
        characteristics of both objects to be quite similar (Munari et al.
        2002).  V838 Mon, iconized by its remarkable light-echo (Bond et
        al. 2003), has been at the center of several, recent and ongoing
        studies. There is a general consensus that V4332 Sgr, V838 Mon and
        M31 RV (a red variable that erupted in M31; Rich et al. 1989) may
        belong to a new and select class  of eruptive variables whose
        post-outburst behavior is different on many counts from other
        classes of eruptive variables ( Munari et al. 2002; Banerjee $\&$
        Ashok, 2002). The fundamental question regarding the cause of the
        explosion  in such eruptive variables or quasi-novae is not
        completely understood. Plausible explanations for the outburst
        invoke merger of stars (Soker and Tylenda 2003 ) and a star
        capturing its encircling planets (Retter and Marom 2003).  The
        cause of the outburst is still uncertain and there is a definite
        need to study these objects further to gain a  better
        understanding of their nature and evolution.\\
	      
               In a recent work (Banerjee et al. 2003; hereafter BVAL) we
        had shown that V4332 Sgr  has a very interesting near-IR spectrum
        at present. Several new bands of AlO were detected and it was also
        shown that the spectral energy distribution (SED) of the object
        had undergone a striking change - a new dust shell having formed
        recently. The present optical study shows that V4332 Sgr is rising
        like a Phoenix  from the ashes. Our results, discussed below, show
        that there is a variety and richness in its spectrum that is
        rarely encountered in astronomical sources.

        \section{Observations}
        The optical observations were made on 29 September 2003 using the
        recently commissioned 2m Himalaya Chandra Telescope (HCT) located
        at Hanle, India. Spectroscopy and photometry were done using HFOSC
        (Himalaya Faint Object Spectrograph Camera) which uses a
        liquid-nitrogen cooled 2048X4096 pixel CCD with 15 micron square
        pixels as the detector. It has a 10$\arcmin$x10$\arcmin$
        unvignetted field in imaging mode and also has several grisms for
        low and intermediate resolution spectroscopy. The spectra
        presented here were obtained at R $\sim$ 870 using a slit
        1.3$\arcsec$ wide. The spectrum of V4332 Sgr and the standard star
        Feige 110 were obtained with exposure times of 15 and 10 minutes
        respectively. The standard star spectrum was used to ratio the
        V4332 Sgr spectrum to remove telluric/air glow lines. The ratioed
        spectrum was finally multiplied by a smooth polynomial fit to the
        flux-calibrated spectrum of Feige 110 as given in Massey et al.
        (1988) - this gives the proper slope to the continuum of the V4332
        Sgr spectrum. All spectra were wavelength calibrated using a Fe-Ar
        spectral lamp.

               Photometry in the $BVRI$ bands, using HFOSC, was done by
        taking multiple exposures in the $BVRI$ bands. Photometric
        calibration was done by observing the Landolt standard star SA110
        -232 (Landolt 1992). A cross-check for the calibration was also
        done by calculating the magnitudes of another comparison star
        SA110 -230. There is a good agreement ($\pm$ 0.05 mag) between the
        derived and listed magnitudes of the comparison star.  The mean
        air-mass at the time of observations was 1.40 and 2.3 for V4332
        Sgr and the standard star respectively. Extinction corrections
        were done using mean values of the extinction coefficient per unit
        airmass of 0.209, 0.121, 0.0823 and 0.0498 magnitudes for the
        $B,V,R$ and $I$ bands respectively for the HCT observatory site.
        The photometric and spectroscopic data were reduced using IRAF.
        The details of the photometry and the derived $BVRI$ magnitudes
        for V4332 Sgr are given in Table 1.

        \section{Results}
        \subsection{Optical spectroscopy}
        The spectrum of V4332 Sgr covering the 5000-8300{\AA} spectral
        range is shown in Figure 1.  The identification of the lines and
        the equivalent widths of the atomic lines are given in Table 2.
        Some unidentified features are also listed with their observed
        wavelengths and equivalent widths.  As can be seen, the striking
        feature of the spectrum is the great strength of the KI resonance
        doublet at 7665 and 7699{\AA} . The blend of the NaI doublet at
        5890 and 5896{\AA}, although unresolved in the spectrum, is also
        very prominent.  We have surveyed the literature, as
        comprehensively as possible, for reported detections of KI and NaI
        lines in emission. Although their occurrence is not too common (KI
        detection is much rarer than NaI) they have been seen in a variety
        of objects viz. comets, Io, Jupiter, the Moon, RCB stars at
        minimum, a few Be stars like HD 45677, MWC 645 and P-Cygni, around
        red giants like Betelgeuse and around some N type stars.
Among these,
        RCB stars in particular are seen to display strong  NaI lines during 
        their minima (Rao et al. 1999) and KI to a much weaker extent.
        In general, we find that the observed strength of the KI doublet
        in V4332 Sgr  is unusual and it  could be  among the strongest to be 
        detected in an  astronomical source.  The presence
of the RbI lines in 
        emission  also appears to be rare.
        Although the RbI lines are  blended with the neighboring 
        TiO$\gamma$(3,4) band,
        we feel their identification is correct because of  the good 
        match between their observed and laboratory wavelengths.  Several of
        the TiO bands identified here,
 are also seen in emission in the 
        peculiar red giant star U Equulei
        (Barnbaum, Omont, \& Morris 1996).

        We find that the average full width at half maximum (FWHM)
        of the KI doublet is 
        greater than that of the instrument profile as obtained from two
        spectral lamp lines in the same wavelength region. 
        Since gaussian fits to the KI lines and the instrument profile
        are found to give a good agreement, the intrinsic width of the KI 
        lines can be
obtained from
        
        \begin{equation}
        FWHM^{2}_{intrinsic} = FWHM^{2}_{obs.} - FWHM^{2}_{instr.}
        \end{equation}

               where the subscripts refer to the intrinsic, observed and
        instrumental widths. For the observed values of $FWHM$${_{\rm obs.}}$
        of 10.6{\AA} and $FWHM$${_{\rm instr.}}$ equal to 8.24{\AA} , we find 
        that
        $FWHM$${_{\rm intrinsic}}$ is $\sim$ 6.6{\AA} or 260 km/s.  The
        expected line width due to Doppler broadening of the KI atoms,
        having a mass $m$${_{\rm KI}}$, a kinetic temperature $T$ and a
        turbulent velocity $V$${_{\rm t}}$ (assumed to be gaussian ) is
        given by 

	\begin{equation}
 	\Delta\nu_D = \frac{1}{\lambda}\left( \frac{2kT}{m_{KI}} + 
 	 {V_t}^2 \right)^{1/2}
 	\end{equation} 
 
               From equation 2 it is seen that thermal broadening will
        account for a negligible amount of the observed line width of
        6.6{\AA} ( for e.g at $T$ = 1000K, the thermal broadening for the
        KI line is only 1 km/s). Thus there is a large amount of velocity
        dispersion in the KI emitting gas. A significant part of this
        broadening could be due to line-of-sight averaging of different
        velocity components in the KI shell in case it has an expansion or
        rotation velocity associated with it. This aspect is discussed
        later in Section 3.3.
        
               The KI doublet lines are expected to have a strength of 2:1
        in case they are optically thin. However, the observed ratio is
        closer to unity for the KI doublet (1.1:1) indicating the lines
        are optically thick. Following Williams (1994), the optical depth
        $\tau$ in the KI 7665{\AA} line can be calculated from
      
	\begin{equation}
	\frac{I_{7665}}{I_{7699}} = \frac{(1 - e^{-\tau})}{(1 - e^{{-\tau}/2})}
        \end{equation}
          
               where $I$$_{\rm 7665}$ and $I$$_{\rm 7699}$ are the
        observed intensities.  Considering the observed equivalent widths
        of 193 and 176{\AA} for the KI lines (Table 2) to represent their
        intensities, we get a value of $\tau$ $\sim$ 4.5 from equation 3.

               If the KI emission arises from a column of length $R$, the
        column density can then be obtained from the relation
 
     \begin{equation}
     \tau = N_{KI}{\frac{{\sqrt{{\pi}}{e^2}}}{m_ec}}{ }{\frac{f}{\Delta\nu_D}}R
     \end{equation}
 
               where $N$${_{\rm KI}}$ is the number density of the KI
        atoms, $f$ is the oscillator strength for the 7665{\AA} transition
        ($f$=0.335) and $\Delta$$\nu$${_{\rm D}}$ is the local line width
        given by equation 2. Using values of $\tau$ = 4.5 and
        $\Delta$$\nu$${_{\rm D}}$ = 3.4x$10$${^{\rm 11}}$ s${^{\rm -1}}$
        (corresponding to a intrinsic width of 6.6{\AA}), we derive a
        column density $N$${_{\rm KI}}$$R$ = 3x$10$${^{\rm 14}}$ cm${^{\rm
        -2}}$.
The value of $N$${_{\rm KI}}$$R$ can be used to calculate the 
        mass of the KI  region (Williams 1994) in case its geometry is better 
        established in any future study - at present we are not too sure
        of the geometry as discussed in section 3.3.

	\subsection{ Photometry and Spectral Energy Distribution }
       
               The SED of V4332 Sgr is shown in Figure 2.  The $BVRI$
        fluxes have been computed after reddening corrections adopting
        $E(B-V)$= 0.32 from Martini et al.(1999) and using zero magnitude
        fluxes from Bessell, Castelli and Plez (1998).  The $JHK$ fluxes
        from BVAL are also plotted to show the SED over an extended
        wavelength range . The present $BVRI$ data establishes more
        definitely - as suggested in BVAL - that the hot component of
        V4332 Sgr is well fit by a blackbody of 3250K. The newly-formed
        dust component (BVAL) is also well-fitted by a 900K blackbody.
        Associating the 3250K component with the central star of V4332 Sgr
        shows that it has an effective temperature corresponding to M5
        type. Its luminosity class is uncertain because of distance
        uncertainties but as per our earlier estimate it is slightly
        over-luminous for a main sequence M5 object.  The SED also
        suggests that the central star has remained at a constant
        temperature of 3250K between the 1998 2MASS observations (BVAL)
        and now. If this is the quiescent state of the star, then it
        appears that the 1994 explosion has taken place on a M type star.
        This gives a more definite classification, not available before,
        on the likely nature of the progenitor on which the outburst has
        taken place. This should be an useful input for models
        investigating the cause of the outburst in quasi-novae.  
        It is difficult to conclude, based on the present data, whether 
        V4332 Sgr is a binary system.  The emission lines are in
        general found to be blue shifted by $\sim$ 3{\AA} - a point
        already noticed in the high-resolution H$\alpha$ line profiles of
        V4332 Sgr obtained during outburst by Martini et al.  (1999). The
        similar observed blueshift of the lines at two different epochs 
        suggests the blueshift of the lines could be due to systemic motion.
        Radial velocity monitoring at higher spectral resolution (than in the 
        present studies) would be helpful in establishing or ruling out 
        binarity for V4332 Sgr. It may be  pointed out that  a hot B3V 
        companion to the outbursting star has   been reported in V838 Mon 
        (Wagner et al. 2003).

        Considering the large strength of the alkali metal lines in V4332
        Sgr, we have investigated whether the M5 central object is some
        variant of a brown dwarf or a very low mass star.  Brown dwarfs
        and very low mass stars show, like V4332 Sgr, very strong
        resonance lines of the alkali metals - although in absorption (for
        e.g. Burgasser et al. 2003 and references therein).  In these
        objects, the NaI and KI lines are the strongest while RbI and CsI
        are progressively weaker (RbI lines are present in our spectrum;
        the CsI doublet at 8521, 8943{\AA} is not covered in our
        spectrum).   Further
evidence that V4332 Sgr type of objects 
        could be related to very
        cool dwarfs comes from the near-IR spectra of V838 Mon which
        indicate that it could be a L giant (Evans et al. 2003).  We
        therefore compared the colors of V4332 Sgr with a large sample of
        L and T type dwarfs whose $BVRIJHK$ magnitudes are available (Dahn
        et al. 2002).  However, our results do not show V4332 Sgr to have
        the colors of a L or T type dwarf. The $BVRI$ colors of the
        suggested M5 central star are not red enough. In the near-IR, the
        comparison is made difficult by the contribution of the dust shell
        in V4332 Sgr to the $JHK$ colors.  While flaring activity on a brown 
        dwarf - over a
 period of few hours - has been reported (e.g. 
        Rutledge et al. 2002), no instance is known of a nova-like eruption 
        in them.

               \subsection{Origin and Excitation of the KI emission }
        Since the KI/NaI lines are seen in emission, and not in absorption,
        they cannot be of
photospheric origin but should  originate from an 
        extended envelope. It is likely that some of the emission in the 
        KI/NaI lines could be caused by resonance scattering from the 
        continuum of the 3250K central source. In addition collisional 
        excitation could also contribute to  the observed line strength.
        This is because  the resonance
lines have low excitation energies
        viz. they are at 2.1ev, 1.61ev and 1.56ev  above the ground level
        for NaI, KI and RbI respectively. Collisions in a low temperature gas
        could excite the atoms to the upper
levels.  Further, the work 
        of Tsuji (1973) on expected molecular
        abundances of different species in stellar atmospheres shows that
        the alkali metals do not associate themselves into any molecular
        form at low temperatures in the range 1000-1500K. Thus the
        availability of large number of neutral atoms with  low
        energy for their excitation could be responsible for the
        strong  KI and NaI lines. However, we have presented here only
        a qualitative discussion and an estimate of the fractional
        contribution of collisional excitation vis-a-vis resonance scattering 
        to the observed line strengths needs a detailed analysis.
           
               As the KI/NaI lines are optically thick, yet do not show
        P-Cygni profiles, they are unlikely to originate in a stellar
        wind. Two possibililities for the site of the KI emission are i) a
        disc around the central source or ii) the ejecta of the 1994
        outburst.  A point in support of the latter case is that NaI was
        seen in emission in the ejecta during the 1994 outburst (Martini
        et al 1999). However, the 1994 ejecta is expected to have a
        spatially resolved diameter of 2$\arcsec$  to 3$\arcsec$ based on the
        well-estimated expansion velocity of 200 - 300 km/s for the ejecta
        and an adopted distance estimate of 300pc to the object (Martini
        et al. 1999). The present optical spectrum, suggests weakly that
        there could be extended emission along the slit at the positions
        of the KI and NaI lines. A high spatial resolution image of V4332
        Sgr would be invaluable to look for extended emission zone around
        the object. It may be pointed out that the observed broadening of
        the KI lines could be accounted for by an expanding nova shell.

               However, we favor the possibility of the KI gas being in an
        extended disc around V4332 Sgr. In a recent observation from
        UKIRT, we have detected the $^{12}$CO fundamental band at 4.67
        ${\rm{\mu}}$m strongly in emission ( again a rare phenomenon) and
        also find a deep water ice absorption band at 3.1 microns (paper
        under preparation).  The co-existence of several species,
        requiring successively cooler temperature conditions, in the same
        object viz. neutral species like KI/NaI, molecular species as AlO,
        TiO and CO and finally a cold, solid-phase water ice component
        suggests a spatial stratification of the different species.
        Therefore, a possible and simplistic scenario for the geometry of
        V4332 could be a central source at 3250K surrounded by a dust
        shell. The dust shell is either clumpy or optically thin since
        radiation from the central source is seen in the SED. Surrounding
        the central source is a disc with the atomic and molecular species
        in the inner parts and ice in the colder, outer regions. If the
        gas in the disc has a Keplerian velocity ( which would typically
        be $\sim$ 150 km/s for a M5 star at a distance of 5-10$R$${_{\rm *}}$),
        the emission lines from the disc could get broadened substantially
        due to rotational motion. This could account for the observed
        width of the KI lines in V4332 Sgr. Incidentally the presence of a
        disc, if it has planetary bodies in it, creates the necessary
        background for the outburst mechanism of a star capturing its
        planets - as suggested by Retter and Marom (2003)- to become
        viable.

               This work highlights the detection of strong emission lines
        of alkali metals in V4332 Sgr.  The
optical spectrum strengthens the 
        idea - since nothing similar to
 it has been found earlier in a 
        nova - that V4332 Sgr belongs to a
 new class of eruptive variables.
        V4332 Sgr is found to be an
  extremely interesting object, worthy 
        of wider attention and
 studies.
        
 	\acknowledgements
        The research work at PRL is funded by the Department of Space,
        Government of India. We thank the staff of HCT, Hanle and CREST,
        Hosakote, that made these observations possible. We are grateful
        to the referee, M. Della Valle, for his helpful comments.

%\clearpage


\clearpage 
\begin{figure}
\plotone{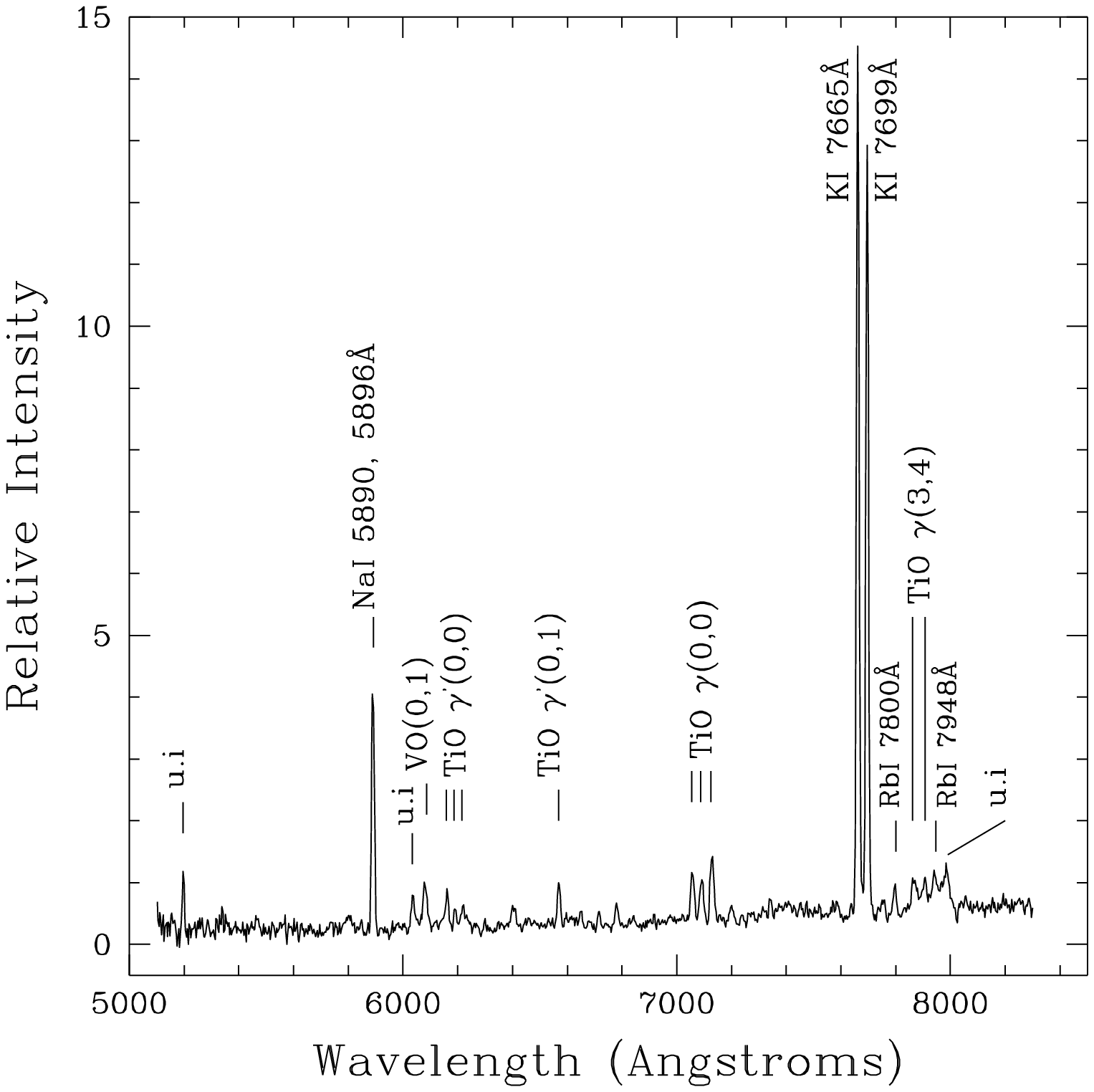}
\caption{The observed optical spectra of V4332 Sgr showing the unusually
strong KI resonance doublet at 7665 and 7699{\AA} and the unresolved
NaI resonance doublet at 5890 and 5896{\AA}. The identification of 
the other prominent lines, as given in Table 1, are marked (u.i means unidentified). \label{fig1}}
\end{figure}

\clearpage
\begin{figure}
\plotone{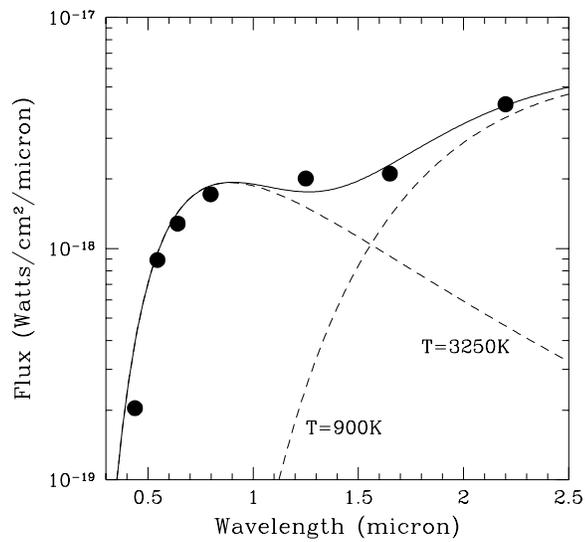}
\caption[]{The SED of V4332 Sgr covering the optical and near-IR range. It
is found to be well  fitted (bold line) by the sum of 2 black-bodies (dashed 
lines) viz. a hot component at T = 3250K and a colder dust component at 
900K.\label{fig2} }
\end{figure}

%% Tables should be submitted one per page, so put a \clearpage before

%% Two options are available to the author for producing tables:  the
%% deluxetable environment provided by the AASTeX package or the LaTeX
%% table environment.  Use of deluxetable is preferred.
%%

%% Three table samples follow, two marked up in the deluxetable environment,
%% one marked up as a LaTeX table.

%% In this first example, note that the \tabletypesize{}
%% command has been used to reduce the font size of the table.
%% Note also that the \label command needs to be placed 
%% inside the \tablecaption.\end{table}

\clearpage
\begin{table}
\caption{Log of photometric observations for 29 Sept. 2003}
\begin{tabular}{lllllll}
\hline \\ 
UT    & Band & Exposure & Integration & Mag. (error)\\
          &      & Time(s)  & Time(s)     &             \\
\hline 
\hline 
12.199 & B   & 50  & 250  & 20.04 (0.15) \\
12.017 & V   & 10  & 90   & 17.52 (0.14)  \\
11.955 & R   & 4   & 80   & 16.31 (0.01) \\
11.955 & I   & 4   & 80   & 15.01 (0.03) \\
\hline
\end{tabular} 
\end{table}

\clearpage
\begin{table}
\caption{A list of the observed  lines in V4332 Sgr. Unidentified lines
are marked as u.i and uncertain identifications with a question mark.}
\begin{tabular}{llllll}
\hline \\ 
S.No & Rest Wave-& Species& Eq. Width\\
    &  length({\AA}) &   &  ({\AA})\\

\hline 
\hline \\ 
1  & 5197    & u.i                         &        40  \\
2  & 5890,96 & NaI                         &        210 \\
3  & 6035    & u.i                         &        15  \\
4  & 6086.4  & VO (0,1)?                   &            \\
5  & 6159    & TiO $\gamma$$\arcmin$(0,0)  &            \\
   & 6187    & "                           &            \\
   & 6215    & "                           &            \\
6  & 6403    & u.i                         &        16  \\
7  & 6569    & TiO $\gamma$$\arcmin$(0,1)? &            \\
8  & 6651.5  & TiO $\gamma$$\arcmin$(1,0)  &            \\
   & 6681.1  & "                           &            \\
   & 6714.4  & "                           &            \\
9  & 6780    & u.i                         &        15  \\
10 & 6843    & u.i                         &        4   \\ 
11 & 7054.5  & TiO $\gamma$(0,0)           &            \\
   & 7087.9  & "                           &            \\
   & 7125.6  & "                           &            \\
12 & 7197.7  & TiO $\gamma$$\arcmin$(1,1)  &            \\
13 & 7664.9  & KI                          &        193 \\
   & 7698.96 & KI                          &        176 \\
14 & 7800.3  & RbI                         &        9   \\
15 & 7861    & TiO $\gamma$(3,4)           &            \\
   & 7907.3  & "                           &            \\
13 & 7947.6  & RbI                         &        5   \\
14 & 7987    & u.i                         &        6   \\
  
\hline
\end{tabular} 
\end{table}


\begin{thebibliography}{}


\bibitem[ref81]{r81} Banerjee, D.P.K., $\&$  Ashok, N.M.,
2002, A$\&$A, 395, 161   
    
\bibitem[ref1]{r1}
 Banerjee, D.P.K., Varricatt, W.P., Ashok, N.M.,  $\&$   Launila, O.,
 2003, \apj, 598, L31-L34 (BVAL)



\bibitem[ref2]{r2}  
Barnbaum, C., Omont, A., $\&$ Morris, M.,
1996, A$\&$A, 310, 250

\bibitem[ref3]{r3}  
Bessell, M.S., Castelli, F., $\&$ Plez, B., 
1998, A$\&$A, 333, 231


\bibitem[ref4]{r4}
Bond, H.E., Henden, A., Levay, Z.G., Panagia, N., Sparks, W.B., Starrfield, S.,
Wagner, R.M., Corradi, R.L.M., $\&$ Munari, U., 2003, Nature, 422, 405


\bibitem[ref5]{r5}
Burgasser, A.J., Kirkpatrick, J.D., Liebert, J., $\&$ Burrows, A.,
 2003, \apj, 594, 510


\bibitem[ref6]{r6}
Dahn, C.C., Harris, H.C., $\&$  Vrba, F.J. et al., 2002, \aj, 124, 1170

\bibitem[ref20]{r20}
Evans, E., Geballe, T.R., Rushton, M.T., Smalley, B., Th. van Loon, J.,
Eyres, S.P.S., $\&$ Tyne, V.H., 2003, \mnras, 343, 1054

\bibitem[ref7]{r7}
Landolt, A.U., 1992, \aj, 104, 340

\bibitem[ref8]{r8}
Martini, P., Wagner, R.M., Tomaney, A., Rich, R.M., Della Valle, M.,
$\&$ Hauschildt, P.H., 1999, \aj, 118, 1034

\bibitem[ref9]{r9}
Massey, P., Strobel, K., Barnes, J.V., $\&$ Anderson, E., 1988, \apj, 328, 315


\bibitem[ref10]{r10}
Munari, U., Henden, A., Kiyota, S., Laney, D., Marang, F., Zwitter, T.,
Corradi, R.L.M., Desidera, S., Marrese, P., Giro, E., Boschi, F., $\&$
Schwartz, M.B.,  2002, A$\&$A, 389L, 51 

\bibitem[ref22]{r22}
Rao, N.K., Lambert, D.L., Adams, M.T., et al.,
  1989, \mnras, 310, 717

\bibitem[ref11]{r11}
Retter, A., $\&$ Marom, A.,  2003, \mnras, 345, L25-L28 


\bibitem[ref12]{r12}
Rich, R.M., Mould, J., Picard, A., Frogel, J.A., $\&$ Davies, R.,
  1989, \apj, 341, L51
  

  
\bibitem[ref13]{r13}  
Rutledge, R.E., Basri, G., Martin, E.L., Bildsten, L.,
  2000, \apj, 538, L141-L144 

\bibitem[ref14]{r14}
Soker, N., $\&$ Tylenda, R., 
  2003, \apj, 582, L105


\bibitem[ref15]{r15}
Tsuji, T.,
  1973, A$\&$A, 23, 411
  
\bibitem[ref16]{r17}
Wagner, R.M., Schwarz, G., Starrfield, S., Munari, U., Giro, E., Siviero, A.,
Szkody, P., $\&$ Bond, H., 2003, IAU Circ. 8202  

\bibitem[ref19]{r18}
Williams, R.E., 1994, \apj, 426, 279






\end{thebibliography}
\end{document}